**Re-analysis of Phosphine in Venus' Clouds.**

**Jane S. Greaves, Anita M. S. Richards, William Bains, Paul B. Rimmer, David L. Clements, Sara Seager, Janusz J. Petkowski, Clara Sousa-Silva, Sukrit Ranjan, Helen J. Fraser**, on behalf of the authors of "Phosphine in the Cloud Decks of Venus" (Nature Astronomy, 14 September 2020; https://doi.org/10.1038/s41550-020-1174-4)

(Response to the Matters Arising: "No phosphine in the atmosphere of Venus" by G. Villanueva et al., preprint at arXiv:2010.14305.)

*Abstract:* We first respond to two points raised by Villanueva et al. We show the JCMT discovery spectrum of $PH_3$ can not be re-attributed to $SO_2$, as the line width is larger than observed for $SO_2$ features, and the required abundance would be an extreme outlier. The JCMT spectrum is also consistent with our simple model, constant $PH_3$-abundance with altitude, with no discrepancy in line profile (within data limits); reconciliation with a full photochemical model is the subject of future work.

Section 2 presents initial results from re-processed ALMA data. Villanueva et al. noted an issue with bandpass calibration. They have worked on a partially re-processed subset of the ALMA data, so we note where their conclusions, and those of Greaves et al., are now superseded.

To summarise: we recover $PH_3$ in Venus' atmosphere with ALMA (~5σ confidence). Localised abundance appears to peak at ~5-10 parts-per-billion (ppb), with suggestions of spatial variation. Advanced data-products suggest a planet-averaged $PH_3$ abundance ~1-4 ppb, lower than from the earlier ALMA processing (which indicated 7+ ppb).

The ALMA data are reconcilable with the JCMT detection (~20 ppb) if there is order-of-magnitude temporal variation; more advanced processing of the JCMT data is underway to check methods. Independent $PH_3$ measurements suggest possible altitude dependence (under ~5 ppb at 60+ km, up to ~100 ppb at 50+ km; see Section 2: Conclusions.)

Given that both ALMA and JCMT were working at the limit of observatory capabilities, new spectra should be obtained. The ALMA data in-hand are no longer limited by calibration, but spectral ripples still exist, probably due to size and brightness of Venus in relation to the primary beam. Further, spatial ripples are present, potentially reducing significance of real narrow spectral features.

*Section 1: Response to comments*

We refer from here on to our paper in Nature Astronomy as **G2020**, and to the Matters Arising submitted by Villanueva et al. as **V2020**.

*A) Separation of spectral features.*

V2020 argued that the $PH_3$ 1-0 transition at 266.9445 GHz is too close to an $SO_2$ transition (offset by +1.3 km/s) to permit spectroscopic separation, when the line-cores are several km/s wide and the JCMT and ALMA datasets were presented by G2020 at spectral resolution comparable to the velocity separation.

$SO_2$ contamination was already considered in detail by G2020 (see Methods and Figure 4). Because the two transitions are separated by 1.3 km/s but both lines are expected to be a few km/s wide, they will not appear as two separate minima at *any* spectral resolution. The limit in identification here is the precision of the centroid of the line minimum (calculated in Table 1 of G2020). The intrinsic spectral resolutions of the datasets are 0.034 km/s (JCMT) and 0.069 km/s (ALMA), and the centroids are measured to precision as good as 0.3 km/s. The spectra were shown in G2020 with larger velocity-bins for clarity, but this does not affect the precision.

*B) Radiative transfer predictions for line profiles.*

V2020 preferred a different line-broadening coefficient for $PH_3$, which can reduce the inferred abundance, and argued that, after removal of some $SO_2$ contribution, $PH_3$ was not detected, at or near the level <5 ppb inferred from a TEXES-IRTF infrared spectrum[1].

The modelling processes of V2020 and G2020 are very similar. We describe below our conclusion that the feature identified as $PH_3$ can not be explained as $SO_2$, because of the extreme-outlier $SO_2$-abundance this would require, and the incompatibility of the observed line width.

V2020 propose a broadening coefficient of 0.12 $cm^{-1}$/atm, based on an air linewidth of 0.067 $cm^{-1}$/atm from HITRAN, scaled by a typical 1.8 scaling ratio as observed for $SO_2$ lines. The scaling of a broadening coefficient using a different molecule is common practice, but has limited empirical backing. The value G2020 adopted is not 0.286$cm^{-1}$/atm as stated by V2020. G2020 in fact adopted 0.186 $cm^{-1}$/atm, and was this obtained using $NH_3$ as a proxy – a molecule with much more structural similarities to $PH_3$ than $SO_2$ has. Co-author CSS, who estimated the broadening coefficient in G2020, has published extensively on the calculation of spectroscopic properties for phosphine. G2020 emphasised the ~50% uncertainties in abundance, by comparing coefficients obtained by two methods.

Three of the G2020 authors are co-authors on ref. (1). There we noted that both temporal variability and the altitudes traced by different wavebands need investigation, to reconcile ~20 ppb from JCMT but <5 ppb from TEXES. The TEXES data trace the column above ~60 km while the JCMT data trace molecules above ~55 km, so the former column intrinsically hosts somewhat less phosphine. The $PH_3$ transitions analysed in Encrenaz et al. are rovibrational excitations with significantly different predicted broadening coefficients (quantum number J = 3 rather than 0).

*C) Interferomeric analysis: post-processing*

V2020 noted that analysis of interferometric data is relatively complex, especially for such a bright and extended source as Venus. How such parameters as bandpass calibration are treated can be crucial to the quaility of output spectra. "Fringing patterns" were noted by V2020 in Extended Data Figures 3 and 4 of G2020, with pattern-width comparable to the ±5 km/s intervals used by G2020 in extracting their lines, and V2020 argue that artifical features can then be produced when removing high-order polynomials from residual data.

What V2020 described as "fringing" was in part caused by a pre-processing error that ALMA have now corrected with a revised dataset. Re-processing of the ALMA data is discussed extensively below. Here we discuss V2020's comments on the original data analysis in G2020.

The issue of fitting the spectral bandpass was considered in detail, and is described in the Methods and Supplementary Information (SI) of G2020. "Fringing" is not the usual term in interferometry for bandpass-ripple, so we use the latter term to avoid confusion. De-trending is central to heterodyne spectroscopy – see e.g. ref. (2) for a detailed discussion of artefact-checking. G2020 laid out the logic (see SI) in defining a minimum polynomial order, given specified bandpass and period of instrumental ripple.

Extended Data Figure 4b in G2020 showed that new features were *not* produced in the ALMA spectra by polynomial subtraction (however, there were issues with the bandpass calibration, see Section 2 below). Specifically, in G2020 we applied the same reduction procedures to regions of the passband offset by 400 spectral channels either side of the phosphine's expected location. This produced narrow artefacts spanning only ~2 channels, much less broad than the real line, and comprising only ~18% of the real line's line-integrated signal. Narrow artefacts of this sort are not physically representative of spectral lines from Venus' clouds.

In their point S2, V2020 applied a polynomial fit to their reduction of the ALMA data. They included all the antenna baselines, whereas G2020 omitted baselines that were substantially noisier, of <33m. Figure FS1 of V2020 demonstrates fitting a 12$^{th}$-order polynomial baseline, with the residual creating an artificial "line" feature.

This procedure used by V2020 is not correct in context. G2020 noted the very strong ripple when all the ALMA antenna-baselines were included, and page 3 of the SI describes the decisions made in excluding short baselines (balancing random noise and systematic ripple). When V2020 include all the antenna-baselines, they recover this very strong ripple. It is then inevitable that they can produce a "fake line" by fitting across a section of the passband, ignoring the actual shape of the data.

Further, the correct polynomial order is defined by the number of changes of direction of the spectral baseline within the passband. As this value appears to be 7 in V2020's Figure FS1 (left), the correct order would be 8. By fitting a 12$^{th}$ order polynomial designed for different data, they have given the polynomial function excess freedom, generating an unstable solution.

Fitting polynomials is not a method we would use when the spectrum is dominated by large systematic ripples. These ripples are produced on the short antenna-baselines, and including these also raises the noise substantially.

There seems to be an inconsistency in l:c in V2020's Figure FS1, since the peak-to-peak amplitude of the ripples in the left panel is ~0.15 Jy while the planetary flux at the time was ~2500 Jy (see SI of G2020). Hence the peak-to-peak amplitudes in the right panel should presumably be ~6 10$^{-5}$, not ~3 10$^{-5}$.

*D) Interferometric analysis: calibration*

V2020 noted that there are significant differences in the residual spectra when they enabled the "usescratch=True" setting in CASA's setJy task. This issue was flagged and the JAO put the data into QA3 for re-evaluation. V2020 gave a brief description of processing they performed, employing both the ALMA/CASA pipeline and the AIPS software package. They argued that the AIPS processing yields a flat spectrum within the noise, and upper limits on PH3 of <(1-2) ppb.

V2020 are correct in identifying that the additional parameter in setJy was not enabled. This however is only *one* of the factors the ALMA Observatory have identified as requiring updating in the analysis of this dataset; see below. Other factors have emerged, some of which cannot be neglected. For example, a "zigzag" function has been discovered to be introduced at one spectral-binning stage (normally adopted to improve signal-to-noise in bandpass calibration). This directly affects the extraction of very faint lines in bright sources (requiring high dynamic range). This information was not available to V2020 at the time of their submission of a Matters Arising, but is included below in our discussion of the re-calibrated data.

V2020 were able to identify this issue with our reduction scripts because we uploaded them with G2020. V2020 are encouraged to provide the reduction scripts behind their analysis. Without a more detailed description by V2020, and specifically the script details, we can not comment meaningfully on the differences between their ALMA figures presented here and either the analysis presented in G2020 or the revised analysis presented here.

The AIPS reduction (Figure FS3 of V2020) has produced a much higher noise level than that in the spectra of G2020. It is therefore not surprising that the $PH_3$ line was not recovered after this processing. In Figure 2 of G2020, the peak-to-peak noise amplitude was ~3 $10^{-5}$ in line:continuum, while in the AIPS reduction, Figure FS3, it is much higher, at ~10 $10^{-5}$.

*E) Vertical profiles*

V2020 consider that there is a mismatch between the vertical profile used by G2020 (a column of constant $PH_3$ fractional abundance above ~55 km altitude, below which the atmosphere is opaque at 1 mm wavelength) with G2020's photochemical model predictions. Features arising at a pressure of 1 bar could have linewidths of GHz, while narrow lines as extracted in G2020 are argued to come from layers with pressures <0.02 bar, i.e. altitudes above 70 km.

The very wide parts of the line-profiles noted by V2020 could not be recovered by any millimetre-waveband instrumentation, to our knowledge (exceeding the sensitivity possible with a Fourier transform spectrometer, in particular).

In G2020 we fully described the fitting scheme that of necessity removes the line wings. This is a standard procedure in millimetre heterodyne spectroscopy in the presence of low-level baseline ripple. We also used a wideband setting with ALMA to attempt to recover the line wings, but the quality was not sufficient (extremely high dynamic range would be required).

As only the $PH_3$ line-core is retained, the width (e.g. full-width at half-minimum, FWHM) is always an *under*-estimate. We showed the trends in FWHM for the JCMT spectrum in Table 1 of G2020 – as the interpolation-region of the fitting-polynomial was increased over $|v|$ = 2, 5, 8 km/s, the FWHM increases over 2.8, 3.6, 8.2 km/s. If more of the line wings could be retained, the FWHM would increase further, and so the inferred minimum altitude will decrease.

We now illustrate this in Figure (f1), re-plotting these three reductions from G2020. A 10 ppb radiative transfer model is overlaid (blue curve), *without* removing the modelled line wings. This model has our simplest approach, i.e. constant abundance fraction (but less $PH_3$ at higher altitudes because the atmosphere is thinner). The spectra converge towards the model as $|v|$ increases, as expected. Given that the data shown are compatible with the presence of the model's line wings, there is *no empirical evidence* that $PH_3$ lies only above 70 km.

Figure (f1). JCMT spectrum obtained with interpolation regions ±(2,5,8) km/s (top to bottom curves).

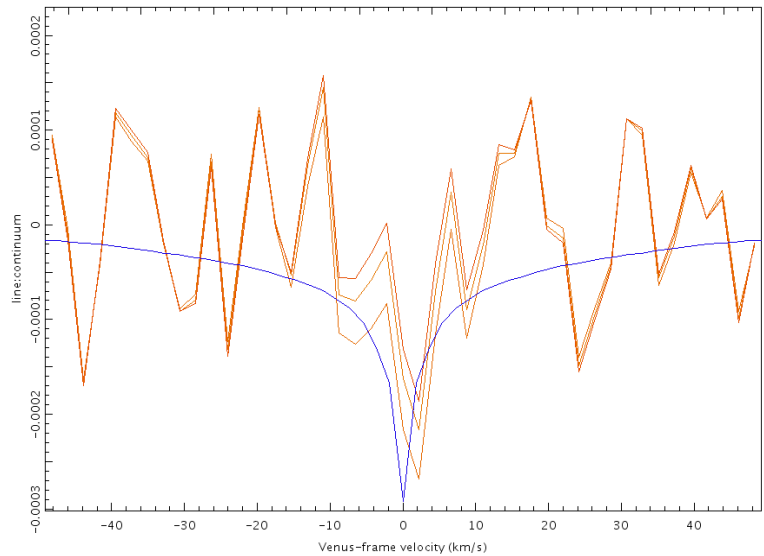

*F) Photochemical model.*

V2020 consider that the photochemical modelling predictions of G2020 do not match the $PH_3$ abundances at the correct altitude, and so significantly higher production rates would be required to match the abundances inferred from observations and radiative transfer modelling by G2020.

The G2020 photochemical model was emphasised to be preliminary (pages 7-11 of the SI gave a detailed description, noting the areas of uncertainty). The intent was to estimate what production rate is needed to produce ~10 ppb of $PH_3$, at any altitude. A paper is in preparation on a more complete model (led by G2020 co-author PBR).

G2020 did not attempt to fold the preliminary photochemical model into the radiative transfer calculation. The photochemical model was mainly used to estimate the fastest destruction lifetime, in order to find possible source rates that could maintain ~10 ppb of $PH_3$.

*G) Contribution of $SO_2$*

V2020 consider that the JCMT feature can be fully reproduced by a mesospheric abundance of ~100 ppb of $SO_2$, reading such a value off the altitudinal molecular-abundances plot shown as Extended Data Figure 9 in G2020.

To address this point, we re-examined the JCMT data. Figure (f2) shows the spectrum obtained by G2020 (in the mid-range |v| = 5 km/s reduction), overlaid with a linearly-scaled version of our radiative transfer model of the proposed $SO_2$ line, after baseline-subtracting this model in the same way performed as on the data. The required $SO_2$ abundance to reproduce the whole feature would in fact need to be ~150 ppb, not 100 ppb.

This value of ~150 ppb would be an extreme outlier in *millimetre-waveband* monitoring observations. (We note some higher literature abundances derived from UV/IR observations; see discussion of data tracing the cloud top[1] and over time[3].). Comparing to a large compilation of millimetre-derived $SO_2$ abundances[4], a value of 150 ppb would be a > +6σ outlier – in fact, the *highest* value recorded over several years was only 76 ppb, half this value. $SO_2$ would also need to be *sustained* at this very high level over the week of the observations, while it is normally seen to vary on timescales of hours to days. Finally, the best-estimate line width, FWHM ≥ 8.2 km/s noted above, is not at all compatible with FWHM ~2-3 km/s observed for in $SO_2$ millimetre-spectra[4,5].

Figure (f2). JCMT spectrum overlaid with 150 ppb $SO_2$ model(black curve).

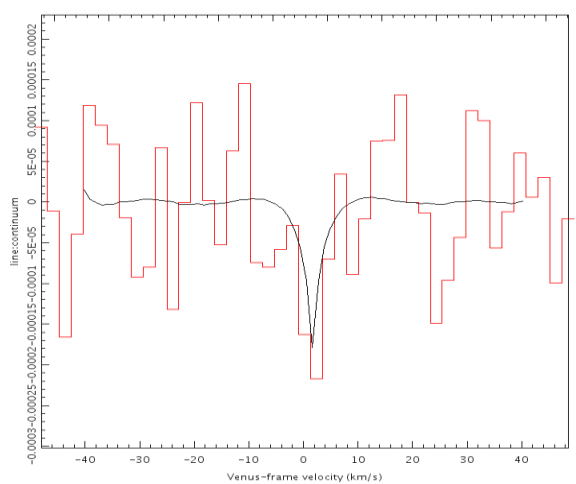

### H) Fitting of JCMT spectrum

In their Figure 1, V2020 re-plot the "most-conservative" spectrum of G2020 (with |v| = 2 km/s interpolation-region), and argue that subtracting model $SO_2$ spectra leaves no significant residual. They then over-plot three alternate $PH_3$ models on this non-significant residual.

Given V2020 are concerned about the line width, their adoption of G2020's most-conservative JCMT spectrum in their Figure 1 is strange, as this reduction intrinsically minimises line width. G2020 presented this particular spectrum only to illustrate minimum signal-to-noise (S/N = 3). Hence, subtraction of almost *any* line profile here will leave a residual consistent with zero.

Extended Data Figure 9 of G2020 was not intended to imply 100 ppb of $SO_2$ (as ascribed by V2020), but to illustrate that measured values are temporally variable (the x-errorbars in this figure were indicative, and plotted the same for all molecules).

In the right panel of V2020's Figure 1, the flat model $PH_3$ line-profile is presumably (this is unstated) only for that part of the column above their hypothetical lower altitude of 70 km. This is not in agreement with the radiative transfer calculation as it was presented by G2020. Additionally, the $PH_3$ 20 ppb models shown are not optimum: the lower (blue) curve uses the doubtful broadening coefficient of 0.12 cm$^{-1}$/atm, and the upper (orange) curve misquotes G2020, who adopted 0.186 cm$^{-1}$/atm not 0.286 cm$^{-1}$/atm.

In their Figure 2, V2020 compared the ALMA whole-planet spectra (those of G2020 and from V2020's new processing) with models of $SO_2$. They contrast their radiative transfer model outputs with a version they generate and label "$SO_2$ Greaves", stating that this uses the T/P profile shown on Extended Data Figure 8 of G2020, and 100 ppb of $SO_2$ as read off by them from Extended Data Figure 9 of G2020.

However, G2020 showed that the deepest possible $SO_2$ line:continuum value is -2 $10^{-5}$ – it can not be as deep as -6 $10^{-5}$ as in the curve representing "$SO_2$ Greaves", generated by V2020. The line-depth was strongly constrained by our simultaneous ALMA observation of another $SO_2$ transition (Figure 4 in G2020), and a radiative transfer model similar to V2020.

V2020 over-plot in their Figure 2 another model labelled "$SO_2$ reduced", with ~10 ppb of $SO_2$ – which is exactly the *same* as the value inferred by G2020 (treated there as an upper limit). This value for the abundance was stated in the caption of Figure 4 of G2020.

$PH_3$ models shown by V2020 in their Figure 2 again use the broadening coefficient from a scaling factor based on $SO_2$, rather than the more analogous $NH_3$. V2020 have also here incorrectly applied 20 ppb to fitting the ALMA spectrum. G2020 noted that the abundance is best inferred from the JCMT data, as the ALMA whole-planet line was ~3 times weaker (i.e. would be interpreted without any spatial-filtering corrections as ~7 ppb).

In G2020, this difference in JCMT/ALMA line-depths was attributed to spatial filtering in the absence of short ALMA baselines, and was quantified on page 4. The new processing (described below) supersedes the discussion in G2020 of $PH_3$ abundance derived from the ALMA data.

*I) Supplementary Material of V2020*

V2020 presented three supplementary paragraphs entitled "S1: Vertical Profiles", "S2: Analysis of the ALMA using G2020 calibration scripts" and "S3: Validation of our ALMA analysis by interpreting other nearby lines".

S1: V2020 argue that mesospheric $SO_2$ abundances during JCMT observations in June 2017 cannot be constrained by $SO_2$ abundances measured using ALMA in March 2019, and then assume ~100 ppb of $SO_2$ to be a plausible and average value for interpreting the JCMT data.

- We see no reason to assume 100 ppb was the value at the time of the JCMT observation, and we showed above that 100 ppb of $SO_2$ would represent an extremely high outlier, let alone the ~150 ppb needed for this interpretation of the JCMT spectrum. The median in millimetre-waveband monitoring[3] is ~18 ppb; G2020 and V2020 above both favour up to ~10 ppb, and detailed modelling[6] of ref. 4's ALMA data gave ~16.5 ppb.

S2: V2020 produced ALMA spectra using the G2020 scripts, but including all antenna baselines, and fitted 12th-order polynomials to generate a "line" from a "harmonic".

- We showed in section (C) that polynomial fitting can not be used when the spectrum is entirely dominated by periodic ripple.

S3: V2020 process ALMA data around other frequencies where $SO_2$ and HDO transitions lie, obtaining ~10 ppb for SO2 and ~60 ppb for $H_2O$, assuming D/H = 200 times the terrestrial value.

- V2020 and G2020 agree on (≤)10 ppb for $SO_2$ seen by ALMA, and plausible $H_2O$ abundances come out of both analyses (see Extended Data Figure 5 in G2020). G2020 inferred ~40-200 ppb of $H_2O$, bracketing V2020's ~60 ppb, using the same reference for deuteration correction. The HDO data from ALMA have also been re-processed and are discussed below.

*Section 1: Conclusion*

V2020 concluded that the detection of $PH_3$ was incorrect, arguing that $SO_2$ could explain the JCMT line; the ALMA reduction was invalid; and there was an altitude discrepancy with the photochemical model.

In summary to these points:

1) we have shown that 100 ppb of $SO_2$ can neither explain the JCMT feature, nor is it a reasonable abundance;

2) the ALMA data have now been re-calibrated and re-processed (Section 2 below);

3) $PH_3$ does not need to be present only at altitudes above 70 km, given the limitations of the data, and the G2020 photochemical network was preliminary.

Therefore, the outstanding issue is the re-processing of the ALMA data – in Section 2 we describe the progress that has been made, and discuss the implications.

*Section 2:* **Results from new ALMA processing.**

*Context*

The G2020 project pushed ALMA's limits in detecting faint absorption of a strong continuum. The observations in March 2019 were highly time-constrained by issues of weather and antenna configuration, so the choice of possible calibrators was limited. Jupiter's moon Callisto was chosen as the brightest available bandpass calibrator. Using an ephemeris object required a full model to be inserted during calibration, but this is not needed for the QSOs normally used and was not standard practice for standard 'QA2' calibration in 2019. This caused the model to be used incorrectly, both in setting the flux scale (too high) and more significantly in introducing severe bandpass ripples. These aspects were not known to the science team up to the point of the G2020 publication.

A critical part of extracting spectra of Venus was a de-trending procedure to characterise ripple in the spectral bandpass, allowing lines to be measured. G2020 used a mathematically-correct polynomial, and showed that this process did not introduce "fake lines" (Extended Data Figure 4). However, this type of de-trending relies on the ripple superimposed on the real absorption line having similar characteristics to ripples elsewhere in the bandpass. This assumption now appears invalid, for two possible reasons:

- in standard calibration the bandpass calibrator data are channel-averaged (here this was over 4 MHz) to improve signal to noise but it has now been shown that this introduces a "zigzag" at low levels in line:continuum (below ~1:1000),

and in addition it is possible that:

- the centre of the passband can show peculiarities if the frequency of interest is tuned to the passband centre in simultaneous narrowband and wideband settings, which was the case for the G2020 data.

The intent is to re-observe Venus with optimum settings after the re-opening of ALMA in 2021. Such observations can include optimising the selection and use of calibrators; using a small mosaic and total power observations if Venus fills the primary beam; and applying a more recent higher-precision primary beam model, needed to accurately extract faint absorption lines across the planet. In the meantime, the data gathered in March 2019 has been reprocessed to resolve some of the processing issues in the data used in G2020.

Data re-processing was developed and led by the ESO ALMA Regional Centre, and products are now available in the ALMA public archive. Other projects affected by the same bandpass calibration issue are also being re-processed and re-released by the ALMA observatory.

*Re-processing*

A full description of the new processing steps is being released by ESO. Along with some minor updates (such as to antenna position tables), the major factors are:

- corrections to the flux scale (previously, the effect of Venus on system temperature was not handled correctly by standard procedures);
- including a model for Callisto (implemented via usescratch=True in the CASA task setJy);
- optimise bandpass correction for Callisto, including using smooth polynomial fitting in the CASA bandpass task.

The re-archived data adopt a bandpass calibration solution that applies a polynomial that is $3^{rd}$-order in amplitude and $5^{th}$-order in phase. Here we demonstrate outputs with two approaches: the simplest is to simply smooth the Callisto bandpass, while a more experimental approach (developed by ALMA staff for this dataset) is to apply higher-order polynomials. This optimised solution for bandpass self-calibration across the entire narrow-band spectral window uses $12^{th}$-order polynomials for amplitude while maintaining $5^{th}$-order for phase. We note that this *supports the subtraction of $12^{th}$-order polynomials* by G2020. That is, the need for this order (or a closely similar one, depending on the exact width of the spectral-window being processed) was correctly identified by G2020, but it has now been applied in the visibility domain.

Overall, all the spectra now have much shallower bandpass ripples, but some spatial gradients and spectral artefacts remains. The latitudinal bands analysed by G2020 suffered from worse spatial gradients, and should be regarded as superseded. Ripples are still worse on shorter baselines, so we have excluded baselines <33m, as in G2020.

The validity of the re-processed data has been checked by inspection of the HDO spectra, observed simultaneously with $PH_3$. In particular, comparisons can be made with HDO lines from Venus previously observed with ALMA by ref. 5.

*New results*

In summary, the re-processed ALMA data show that *$PH_3$ lines are weaker* than reported in G2020, but results are significantly improved. After full reprocessing, the whole-planet spectral noise is now around half that in the original G2020 results, and much lower than in V2020.

After the full re-processing, including the high-order bandpass self-calibration, we have extracted new whole-planet spectra. Figure (f3) shows line:continuum for PH$_3$ (red) and HDO (green; offset for clarity), in our data product using only antenna-baselines >33 m. No post-processing has been performed on these spectra, apart from dividing the line products by 13.77 Jy/beam, the estimated continuum signal (considered to be accurate to ~10%). Overlaid is our PH$_3$ radiative transfer model, scaled to 1 ppb. To simulate in this model the data-processing that may have removed features more than ~20 km/s wide (i.e. the polynomial applied to the bandpass calibrator), we subtracted a 12$^{th}$-order function with interpolation over ±10 km/s around line-centre; this results in ~25% signal-loss at line-minimum in the model.

Figure (f3). New whole-planet spectra from the re-processed ALMA dataset (see text). The upper spectrum (red) is PH$_3$ 1-0; the lower spectrum (green, offset for clarity) is HDO $2_{2,0}$-$3_{1,3}$.

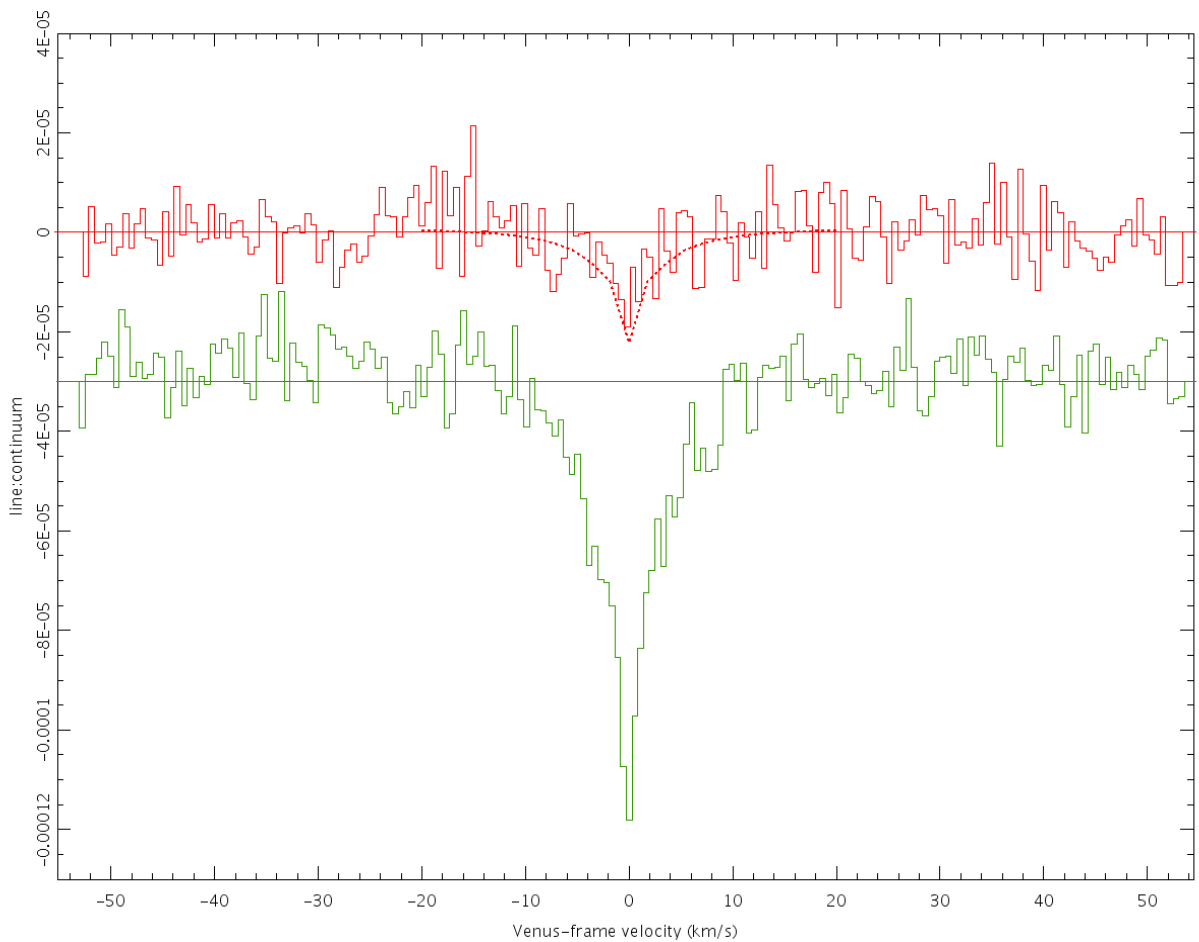

The whole-planet PH$_3$ detection is now weaker but reasonably secure, with a line-integrated signal (over ±8 km/s) of 4.8σ. The centroid over the same velocity-span is at -0.6 ± 0.7 km/s. The inferred PH$_3$ abundance is ~1-4 ppb, with this value depending on the spatial distribution of the molecules. The lower value applies if PH$_3$ is in fully-detected few-arcsec-scale areas, and the higher value includes a spatial-filtering correction (for the loss of the shorter baselines), that assumes the phosphine is as smoothly distributed as the continuum signal (see G2020).

The faint feature is comparable in depth to our original ALMA limit of -2 10$^{-5}$ for the SO$_2$ contaminant line at +1.3 km/s. An initial examination of the re-processed wideband data suggests a new limit of -1 10$^{-5}$ for SO$_2$ contamination. Thus, we emphasise that there could be a contribution from $SO_2$, but the whole-planet feature appears to be too wide to be *solely* $SO_2$. The FWHM is estimated at 4.5 km/s, compared to ~2 km/s for mesospheric $SO_2$ previously observed with ALMA[5]. Also, the ALMA line-centroid here is discrepant with $SO_2$ at the -2.7σ level (with the JCMT line being discrepant by -1.4σ, and similarly wide with FWHM ≥5 km/s).

If our JCMT and ALMA observations had overlapped with the cloud-regions probed by TEXES[5] (but *not* prior millimetre observations), they could in principle trace layers with enough $SO_2$ to produce the absorption depths we see. However, this is hard to motivate for our two particular observations out of years of millimetre-monitoring, and could still not explain the centroid and line-width discrepancies.

The overall reduction procedure appears robust, based on the HDO line (green histogram in Figure (f3)). From radiative transfer as described in G2020 and the deuteration scaling of ref. 5, the $H_2O$ abundance inferred from Figure (f3) is ~0.05-0.2 ppm (depending on assumed spatial scale; the noisier data-products sampling all planetary scales favour the lower end). These values are in the lower-tercile of monitoring observations made at a similar frequency[7].

Spectra for smaller areas of the planet are still very difficult to extract in this processing, due to low-level gradients across the planet. Our *smoothed-bandpass* reduction, again excluding baselines <33m, is therefore also examined here. Residual ripples in this reduction are now found to be much broader than the planetary lines, facilitating extraction of clean spectra. Hence only low-order polynomials have been fitted in a single post-processing step.

The example spectrum in Figure (f4: left panel) was extracted from the region circled on the simultaneous continuum image (f4: right panel – square-root colour scale; annotated with pole-to-pole axis; planet 26% dark, on the right limb). The circle is 4 arcsec in diameter, smaller than the largest angular scale (LAS) of 4.3 arcsec, and so l:c should represent the true value.

Figure (f4). Example $PH_3$ spectrum, from the circled region superimposed on the continuum image.

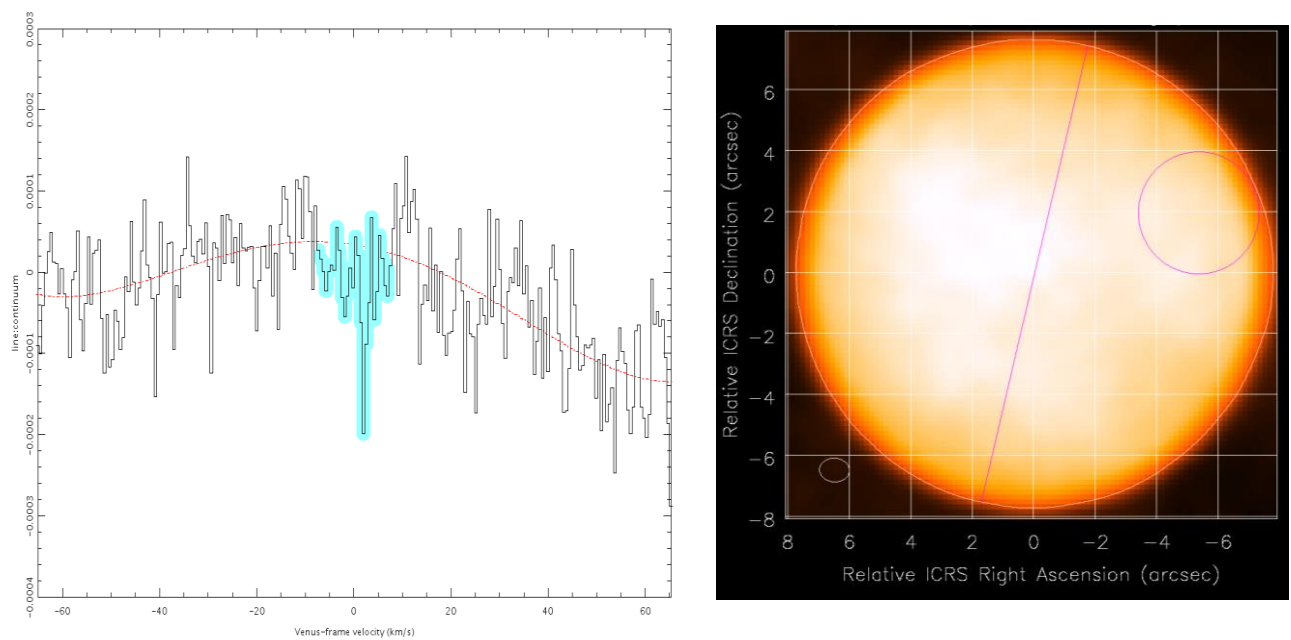

We selected this example where the net spectrum is relatively *clean of artefacts* across the passband – specifically, few offsets in signal-level that persist over a few km/s. This is not the only region with possible $PH_3$ absorption, and we encourage exploration of the dataset.

Figure (f5) shows the same spectrum (in red) after removal of the 4th-order polynomial plotted in Figure (f4). The overlaid curve (in black) is a model profile for 5 ppb of $PH_3$, scaled from our radiative transfer model, and with a spectral baseline subtracted in the same way as for the data, and the upper (grey) histogram shows the residual after subtraction of this model line.

As this extraction-area is smaller than the LAS, ~5 ppb is the best estimate for the phosphine abundance in this region. (In the processing that used higher-order polynomial bandpass calibration, we see candidate features representing abundances up to ~10 ppb, but the spatial filtering corrections are still uncertain.) The line profile in Figure (f5) agrees well with *the expected line shape of an absorption feature*; other 'peaks' in the noise of a few bins width are not physically plausible as lines. The residual after subtraction of the $PH_3$ radiative transfer model is flat, with no significant outliers.

The line-integrated signal-to-noise for this small region is 4.4 (over ±7 km/s). The line centroid is at +0.5 (±0.5) km/s, in good agreement with a $PH_3$ identification. Our preliminary estimate for the maximum $SO_2$ contamination is under $-1 \times 10^{-5}$, much fainter than the line-minimum seen in Figure (f5). Further, the FWHM of an $SO_2$-dominated feature would be expected to be only ~2 km/s (2 spectral bins), narrower than observed.

Figure (f5). As for (f4) but after baseline-fitting; residual offset above; 5 ppb model superimposed.

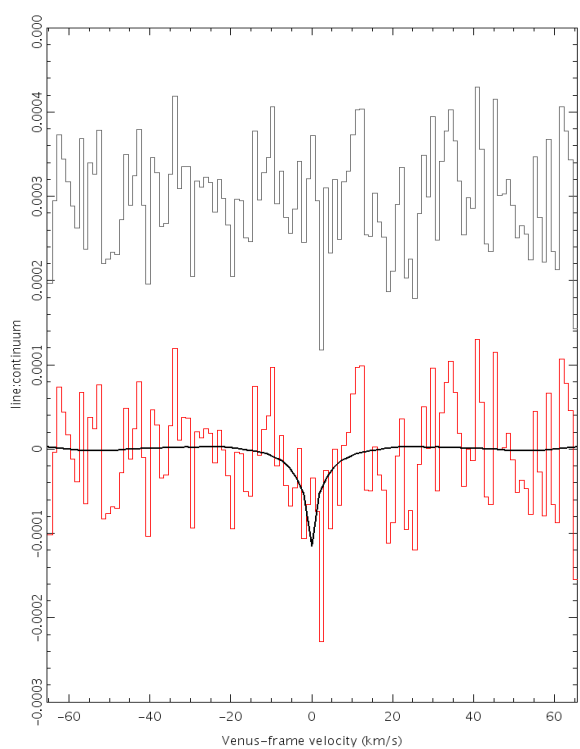

Improved images/spectra will be published (as a Correction to G2020) when we have had more time to work on additional analysis of the QA3 products. We also aim to further investigate the goodness of fit of $PH_3$, $SO_2$ model-profiles, across the full range of spectra that can be extracted in different regions.

*Section 2: Conclusion*

$PH_3$ seen by ALMA is significantly fainter than found by G2020 (in data with calibration issues, now resolved by the observatory). The G2020 line strength would have represented ~7 ppb of phosphine if the distribution was assumed to be patchy (scales less than the LAS of 4.3 arcsec). Here we tentatively identify at least one region on such a scale, where the $PH_3$ abundance is estimated at 5 ppb; we identify no regions above ~10 ppb abundance in the various reduction-outputs. The optimally-processed planet-wide spectrum corresponds to ~1-4 ppb of $PH_3$.

Temporal variation is now required to reconcile all the available $PH_3$ data. G2020 reported a whole-planet detection of ~20 ppb from JCMT – these data are being re-analysed (bypassing the use of polynomial fitting), with initial results again suggesting ~20 ppb abundance. Encrenaz et al. (2020) reported <5 ppb of phosphine in a TEXES 10 μm spectrum obtained in March 2015, for altitudes above ~60 km. A re-assessment of Pioneer-Venus mass spectrometer data[8] provides insights into ~50-60 km altitudes, identifying a mass-peak in agreement with $PH_3$ (plus a detection of $^{31}P^+$ that could only be associated with $PH_3$). Scaling the instrument's count-rate[9] suggests ~100 ppb at the $PH_3$ mass-value, but possibly including some $H_2S$. The overall compilation of data could thus be reconciled with a $PH_3$ profile decreasing with altitude, and temporal variations of around an order of magnitude.

*Acknowledgements*: The G2020 team thank the many ALMA staff who contributed tirelessly and speedily to this re-processing project, developing new tests and techniques in a very short time period. The work was led from ESO with input from JAO and NAASC.

---